\documentclass[amsmath,amssymb,floatfix,showpacs,notitlepage,nofootinbib,
superscriptaddress,
twocolumn
]{revtex4-1}

\usepackage[T1]{fontenc}
\usepackage[utf8]{inputenc}
\usepackage{bm}
\usepackage{hyperref}
\usepackage{graphicx}
\usepackage{color}

\newcommand{\rmi}{\mathrm{i}}
\newcommand{\rmd}{\mathrm{d}}

\newcommand{\ff}{\protect{(\mathrm{f})}}
\newcommand{\mm}{\protect{(\mathrm{m})}}
\newcommand{\nutm}{\tilde\nu}
\newcommand{\Htm}{\widetilde{\mathsf{H}}}

\newcommand{\ket}[1]{|#1\rangle}

\newcommand{\ctm}{\cos(2\thetam)}
\newcommand{\stm}{\sin(2\thetam)}
\newcommand{\ttm}{\tan(2\thetam)}

\newcommand{\caN}{\mathcal{N}}

\newcommand{\wv}{\omega_\mathrm{v}}
\newcommand{\wm}{\omega_\mathrm{m}}

\newcommand{\thetav}{\theta_\mathrm{v}}
\newcommand{\thetam}{\theta_\mathrm{m}}

\newcommand{\tdr}{\frac{\rmd}{\rmd r}}

\newcommand{\sfH}{\mathsf{H}}

\newcommand{\bfB}{\mathbf{B}}

\begin{document}

\title{Matter parametric neutrino flavor transformation through Rabi
  resonances}  
\newcommand*{\UNM}{Department of Physics \& Astronomy, University of New
  Mexico, Albuquerque, NM 87131, USA}
\newcommand*{\LANL}{Theoretical Division, Los Alamos National Laboratory, %
Los Alamos, NM 87545, USA}
\author{Lei Ma
}
\email{leima137@gmail.com}
\affiliation{\UNM}
\author{Shashank Shalgar}
\email{shashankshalgar@gmail.com}
\affiliation{\UNM}
\affiliation{\LANL}
\author{Huaiyu Duan
}
\email{duan@unm.edu}
\affiliation{\UNM}

\date{\today}

\begin{abstract}
We consider the flavor transformation of neutrinos through oscillatory
matter profiles. We show that the neutrino oscillation Hamiltonian in this
case describes a Rabi system with an infinite number of Rabi
modes. We further show that, in a given physics problem, the majority of the
Rabi modes have too small amplitudes to be relevant. We also go beyond the
rotating wave approximation and derive the relative detuning of the
Rabi resonance when multiple Rabi modes with small amplitudes are
present. We provide an explicit criterion of
whether an off-resonance Rabi mode can affect the parametric flavor
transformation of the neutrino.
\end{abstract}


\maketitle

\section{Introduction}
Neutrinos are constantly produced by stars, and they are also emitted much
more intensively during the violent deaths of massive stars through
core-collapse supernovae albeit for only a brief moment. The neutrinos
from stellar objects and other astronomical sources provide a unique
probe to observe these 
objects and to study the properties of the neutrinos themselves (see,
e.g., Refs~\cite{Robertson:2012ib,Mirizzi:2015eza} for reviews on
solar neutrinos and supernova neutrinos). The
interpretation of the neutrino signals from astronomical
sources depends on the understanding of the 
flavor transformation or oscillations of the neutrinos. A well-known
mechanism for a neutrino to experience flavor transformation is
the Mikheyev-Smirnov-Wolfenstein (MSW) effect when the
neutrino propagate through a region where the
matter density varies smoothly across a critical value
\cite{Wolfenstein:1977ue,Mikheyev:1985aa}. Inside the stars and
supernovae the matter densities may have rapid changes and
fluctuations which can also leave important imprints on the passing-through
neutrinos
\cite{Haxton:1990qb,Sawyer:1990tw,Nunokawa:1996qu,Burgess:1996mz,Sawyer:1990tw,Loreti:1995ae,Friedland:2006ta,Fogli:2006xy,Kneller:2010sc,Choubey:2007ga}. 
In an extreme case supernova neutrinos can become completely flavor
depolarized as they traverse the turbulent region behind the supernova shock
\cite{Friedland:2006ta}. 

A matter profile with density fluctuations can cause neutrino flavor
conversion through parametric resonances
even when the matter density never crosses the critical value (see,
e.g., Ref.~\cite{Akhmedov:1999ty} for a review).  For example, a
neutrino can achieve a maximum flavor conversion if the matter density
varies sinusoidally on a length scale which matches that of the neutrino
oscillation in matter with the mean density
\cite{Krastev:1989ix}. Using the Jacobi-Anger expansion and the
rotating wave approximation Kneller et al have shown that a
parametric resonance can also occur when the neutrino oscillation frequency
with the mean matter density matches a harmonic of the spatial
frequency of the sinusoidal matter fluctuation \cite{Kneller:2012id}. This
result has been generalized to the scenarios with matter fluctuations
of multiple Fourier modes \cite{Patton:2013dba},  slowly varying base
profiles \cite{Patton:2014lza} and three-flavor neutrino mixing
\cite{Yang:2015oya}.  

The existence of harmonic parametric resonances is an intriguing
phenomenon, but its 
physical origin is somewhat buried in the mathematical procedure
employed in Ref.~\cite{Kneller:2012id}. It is not entirely clear why the flavor
transformation of the neutrino can be described by only a handful
parametric resonances although there can exist many more such resonances
\cite{Patton:2014lza}. 
There also lacks a criterion of when
the rotating wave approximation fails. 
We intend to address these issues in this short
paper. We will not consider the
collective flavor transformation of the neutrinos due to the neutrino
self-refraction (see, e.g.,
Refs.~\cite{Duan:2010bg,Chakraborty:2016yeg} for reviews on this
interesting subject).

The rest of the paper is organized as follows.
In Sec.~\ref{sec:resonances} we will show that the neutrino
oscillation Hamiltonian with an oscillatory matter profile have an
infinite number of Rabi modes which produce
 the harmonic parametric resonances.
In Sec.~\ref{sec:discussion} we will demonstrate that only a finite
number, usually a small
portion, of the Rabi modes are relevant in a physical situation. 
We will also derive a quantitative criterion of
when an off-resonance Rabi mode may significantly affect the parametric
resonance. 
In Sec.~\ref{sec:conclusions} we will summarize
and conclude our work. 




\section{Rabi resonances in oscillatory matter profiles
\label{sec:resonances}}

\subsection{Equation of motion}

As in Ref.~\cite{Patton:2013dba} we will consider the
mixing between two (effective) neutrino flavors $\nu_e$ and $\nu_x$. The  
flavor wavefunction of the neutrino in flavor basis is 
$\Psi^\ff = [\psi_{\nu_e},\psi_{\nu_x}]^T$, where
$\psi_{\nu_\alpha}=\langle\nu_\alpha|\psi\rangle$ ($\alpha=e,x$) is
the amplitude for the neutrino in state $\ket{\psi}$ to be found in
$\ket{\nu_\alpha}$, and $|\psi_{\nu_e}|^2+|\psi_{\nu_x}|^2=1$.
The flavor evolution of the neutrino in matter is described by the
Schr\"odinger equation
\begin{align}
\rmi \tdr\Psi^\ff(r) 
= \sfH^\ff
\Psi^\ff(r),
\label{eq:eom-flavor}
\end{align}
where the neutrino oscillation Hamiltonian is
\begin{align}
\sfH^\ff=[-\wv\cos(2\thetav)+\lambda(r)]\frac{\sigma_3}{2} +
  \wv\sin(2\thetav)\frac{\sigma_1}{2}.
\label{eq:Hf}
\end{align}
In the above equation,
$\wv$ and $\thetav$ are the oscillation frequency and the mixing angle of the
neutrino in vacuum, respectively, $\sigma_i$ ($i=1,2,3$) are the Pauli
matrices, and
$\lambda(r)=\sqrt2 G_\mathrm{F} n_e(r)$
is the matter potential at a distance $r$ along the neutrino propagation
trajectory, where $G_\mathrm{F}$
is the Fermi coupling constant, and $n_e$ the net electron number density.
In Eq.~\eqref{eq:eom-flavor} we have ignored the trace
term of the Hamiltonian which does not affect neutrino oscillations.
Throughout the paper we 
adopt the natural units with $\hbar=c=1$. 

In this work we will assume a
stationary matter profile of the form
\begin{align}
\lambda(r) = \lambda_0 + \delta\lambda(r),
\end{align}
where $\delta\lambda(r)$ is a small perturbation to the uniform
background matter potential
 $\lambda_0$. As in Refs.~\cite{Kneller:2012id,Patton:2013dba}
we define the background matter basis
\begin{subequations}
\begin{align}
\ket{\nu^\mm_1} &= \cos\thetam\ket{\nu_e} - \sin\thetam\ket{\nu_x}, \\
\ket{\nu^\mm_2} &= \sin\thetam\ket{\nu_e} + \cos\thetam\ket{\nu_x}, 
\end{align}
\end{subequations}
where
\begin{align}
  \thetam=\frac{1}{2}\arctan\left(\frac{\wv\sin(2\thetav)}{\wv\cos(2\thetav)-\lambda_0}\right).
\end{align}
The Hamiltonian in the background matter basis is
\begin{align}
\sfH^\mm=-[\wm-\ctm\delta\lambda]\frac{\sigma_3}{2}
+\stm\delta\lambda\frac{\sigma_1}{2},
\label{eq:Hm}
\end{align}
where   
\begin{align}
\wm = \sqrt{[\wv\cos(2\thetav)-\lambda_0]^2 + [\wv
      \sin(2\thetav)]^2}
\end{align} 
is the neutrino oscillation frequency in matter when $\delta\lambda=0$.

For definiteness we will use
$\sin^2(2\thetav)=0.093$ in all the numerical
examples shown later in the paper. We will also assume that the
background matter density is a quarter of the value of the MSW
resonance, i.e.\  $\lambda_0=\frac{1}{4}\wv\cos(2\thetav)$. These
values and the amplitudes of the matter fluctuations are chosen to
illustrate the general principles to be discussed in this paper and do not
necessarily reflect the actual conditions in real physical problems.

\subsection{Rabi resonance} 
\label{sec:rabi}

We first consider a sinusoidal matter perturbation of
amplitude $\lambda'\ll\wm$ and wave number $k$:
\begin{align}
\delta\lambda(r) = \lambda' \cos(k r).
\label{eq:lambda-single}
\end{align}
Because the fluctuation amplitude is small, we will drop the
perturbation in the diagonal 
terms in Eq.~\eqref{eq:Hm} as a first order approximation so that
\begin{align}
\sfH^\mm 
\approx \frac{1}{2}\begin{bmatrix}
-\wm & \sum_{n=\pm1} A_n e^{\rmi K_n r} \\
\sum_{n=\pm1} A_n e^{-\rmi K_n r} & \wm
\end{bmatrix}
\label{eq:H-rabi}
\end{align}
where 
\begin{subequations}
\begin{align}
K_{\pm1}&= \pm k
\\
\intertext{and}
A_1&=A_{-1} = \frac{\stm \lambda'}{2}.
\label{eq:A1}
\end{align}
\end{subequations}
Eq.~\eqref{eq:H-rabi} has the same form as the equation of motion of a
magnetic dipole in the presence of a magnetic field with two components,
a constant component in the vertical direction and
an oscillating component in the horizontal direction.
The transition
amplitude between the up and down states of the dipole can reach
$100\%$ at the Rabi resonance where $k=\wm$ (see,
e.g., Ref.~\cite{sakurai2011modern}). 

It turns out that the neutrino flavor transformation Hamiltonian with
an oscillatory density profile can always be cast into the form in
Eq.~\eqref{eq:H-rabi}. We will call each term in the sum of the
off-diagonal element in Eq.~\eqref{eq:H-rabi} a ``Rabi mode'' with
$A_n$ and $K_n$ being the amplitude and wave number of the
corresponding Rabi mode. When the Rabi resonance condition
\begin{align}
K_n = \wm
\label{eq:resonance}
\end{align}
is approximately satisfied, the 
transition probability of the neutrino between $\ket{\nu^\mm_1}$ and
$\ket{\nu^\mm_2}$ takes the form  
\begin{align}
P\approx \frac{\sin^2(\Omega r/2)}{1+D_n^2},
\label{eq:P}
\end{align}
where
\begin{align}
D_n = \left|\frac{K_n-\wm}{A_n}\right|
\label{eq:D}
\end{align}
is the relative detuning of the Rabi mode, and
\begin{align}
  \Omega = A_n\sqrt{1+D_n^2}
  \label{eq:Omega}
\end{align}
is the Rabi frequency.

The relative detuning $D_n$
is a measure of how much the corresponding Rabi mode is away from its
resonance. The Rabi mode $n$ is on resonance if $D_n\lesssim 1$ and is off
resonance if $D_n\gg1$. 
Because $D_{-1} > \wm/\lambda' \gg 1$, the $n=-1$ mode is always off resonance
 and is ignored by the rotating wave approximation.

\begin{figure}[!htb]
  \includegraphics[width=\columnwidth]{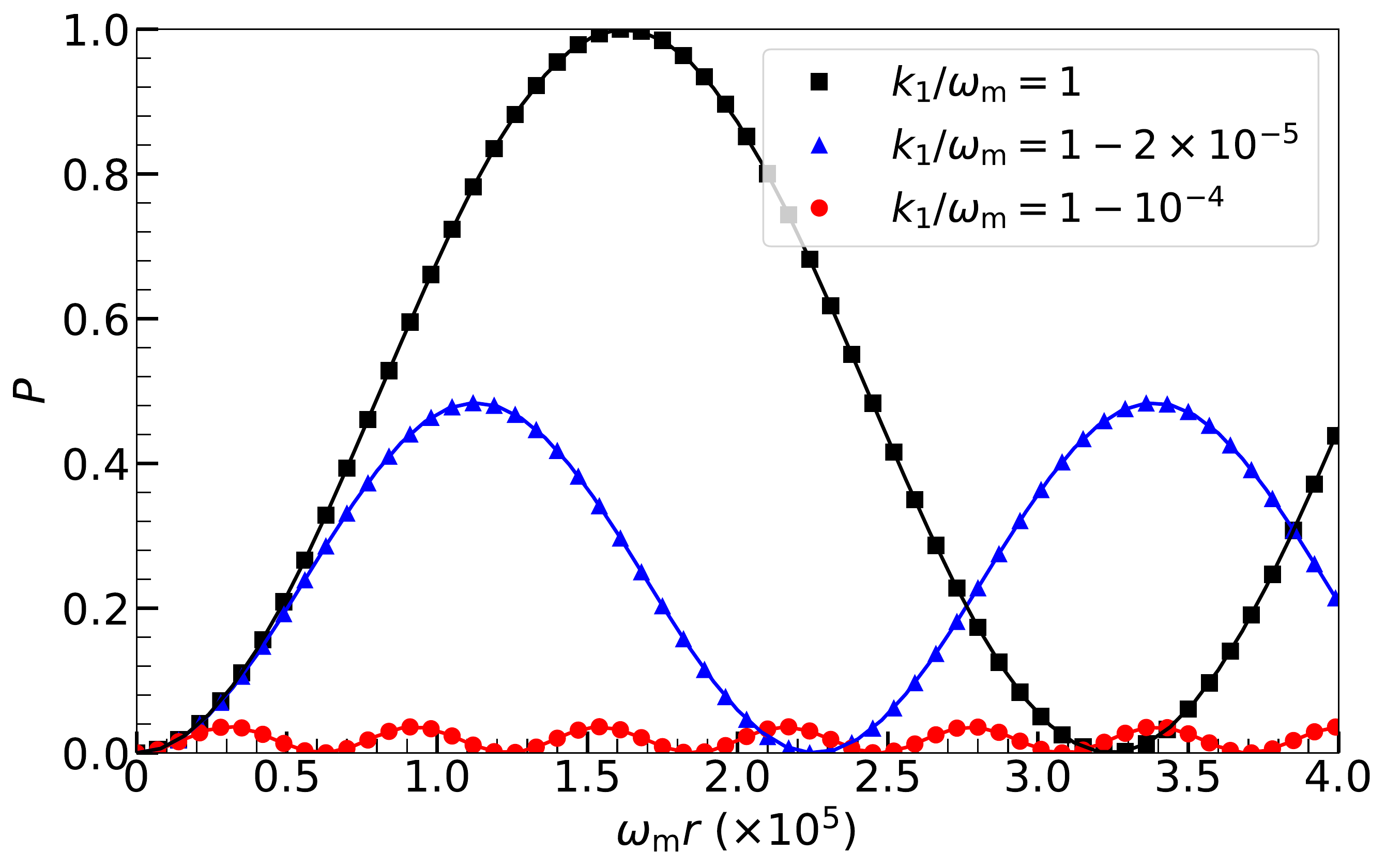}
  \caption{The transition probabilities $P$ between the two background
    matter states 
    $\ket{\nu^\mm_1}$ and $\ket{\nu^\mm_2}$ of the neutrino as
    functions of distance $r$ for three matter profiles all 
     of the form $\lambda(r)=\lambda_0+\lambda'\cos(k r)$. The
    different symbols represent the numerical solutions to the
    Schr\"odinger equation with various values of $k$ as labeled. The
    continuous curves are obtained by using the Rabi formula in
    Eq.~\eqref{eq:P} and ignoring the off-resonance Rabi mode. In all
    three cases $\lambda'/\wm=10^{-4}$.}
  \label{fig:single-mode}
\end{figure}

In Fig.~\ref{fig:single-mode} we compare the numerical solutions to
 the Schr\"odinger equation  and
the results obtained by applying the Rabi formula in
Eq.~\eqref{eq:P} (with $n=1$) for three matter profiles with
sinusoidal fluctuations of various wave numbers. The good agreement
between the two sets of solutions justifies the approximations
that we have made.

\subsection{Jacobi-Anger expansion}
\label{sec:expansion}
The neutrino oscillation Hamiltonian in Eq.~\eqref{eq:Hm} actually
contains an infinite 
number of Rabi modes even for a matter profile with a single Fourier
mode. To see this 
we define a rotated matter basis: 
\begin{align}
\ket{\nutm_1} &= e^{\rmi\eta(r)} \ket{\nu^\mm_1}, &
\ket{\nutm_2} &= e^{-\rmi\eta(r)} \ket{\nu^\mm_2}, 
\end{align}
where
\begin{align}
\eta(r) - \eta(0) = \frac{\ctm}{2}\int^r_0 \delta\lambda(r')\,\rmd r'.
\end{align}
We note that the transition probability between $\ket{\nutm_1}$ and
$\ket{\nutm_2}$ is the same as that between $\ket{\nu^\mm_1}$ and
$\ket{\nu^\mm_2}$. The Hamiltonian in the rotated matter basis is
\begin{align}
\Htm= \frac{1}{2}\begin{bmatrix}
-\wm & \stm\delta\lambda e^{2\rmi\eta} \\
\stm\delta\lambda e^{-2\rmi\eta} & \wm
\end{bmatrix}.
\end{align}

For the sinusoidal matter perturbation
$\delta\lambda=\lambda'\cos(kr)$ we take
$\eta(r)=\ctm\lambda'\sin(kr)/2k$ and
apply the Jacobi-Anger expansion
as in Ref.~\cite{Kneller:2012id}
\begin{align}
e^{\rmi z\sin \xi} = \sum_{n=-\infty}^\infty J_n(z) e^{\rmi n \xi},
\end{align}
where $J_n(z)$ is the $n$th Bessel function of the first kind.
Utilizing the identity
\begin{align}
J_{n-1}(z)+J_{n+1}(z) = \frac{2n}{z}J_n(z)
\end{align}
we obtain
\begin{align}
\lambda'\cos(k r) e^{2\rmi\eta}
&= \frac{\lambda'}{2}(e^{\rmi k r}+e^{-\rmi k r}) \sum_{n=-\infty}^\infty 
J_n(u) e^{\rmi n k r}
\nonumber\\
&=\frac{1}{\ctm} \sum_{n=-\infty}^\infty n k J_n(u)
  e^{\rmi n k r},
\end{align}
where $u=\ctm\lambda'/k$. Therefore, the Hamiltonian in the rotated
matter basis indeed has a form similar to Eq.~\eqref{eq:H-rabi} but
with an infinite number of Rabi modes:
\begin{align}
\Htm = \frac{1}{2}\begin{bmatrix}
-\wm & \sum_n A_n e^{\rmi K_n r} \\
\sum_n A_n e^{-\rmi K_n r} & \wm
\end{bmatrix}
\label{eq:H}
\end{align}
where $n=0,\pm1,\pm2,\cdots$,
\begin{subequations}
\begin{align}
  K_n &= n k,
  \label{eq:K-n}\\
\intertext{and}
A_n &=\ttm K_n  J_n(u).
\end{align}
\end{subequations}
Eqs.~\eqref{eq:resonance} and \eqref{eq:P} show that a parametric
resonance occurs when $\wm$ matches a harmonic of the spatial frequency of
the sinusoidal matter fluctuation.

When the $n=1$ mode is approximately on resonance, 
\begin{align*}
A_1
\xrightarrow{k\approx\wm\gg\lambda'}
\frac{\stm\lambda'}{2}
\end{align*}
which reduces to Eq.~\eqref{eq:A1}. Here
we have used the asymptotic form of the Bessel function
\begin{align}
J_n(z) \xrightarrow{z\ll\sqrt{n+1}} \frac{(z/2)^n}{n!}
\quad\text{if } n>0.
\label{eq:J-asym}
\end{align}
In applying the Rabi formula in Eq.~\eqref{eq:P} one has
assumed the rotating wave approximation and
ignored all the off-resonance Rabi modes.

\subsection{Multiple Fourier modes}
\label{sec:multi-freq}

Now we consider the scenario where the matter fluctuation has multiple
Fourier modes:
\begin{align}
\delta\lambda(r) = \sum_a \lambda_a\cos(k_a r + \phi_a),
\end{align}
where $\lambda_a$, $k_a$ and $\phi_a$ are the amplitude, wave number
and initial phase of the $a$th Fourier mode, respectively.
Using the same technique as that in Sec.~\ref{sec:expansion} one can
show that
\begin{align}
\Htm = 
\frac{1}{2}\begin{bmatrix}
-\wm & \sum_\caN A_\caN e^{\rmi (K_\caN r+\Phi_\caN)} \\
\sum_\caN A_\caN e^{-\rmi (K_\caN r+\Phi_\caN)} & \wm
\end{bmatrix},
\label{eq:H-rot}
\end{align}
where the sum is over all possible choices of
$\caN=\{\cdots,n_a,\cdots\}$ with $n_a$ being an arbitrary integer
associated with the $a$th Fourier mode, and
\begin{subequations}
\begin{align}
K_\caN &= \sum_a n_a k_a, 
\label{eq:K-N}
\\
\Phi_\caN &= \sum_a n_a \phi_a, \\
\intertext{and}
A_\caN &= \ttm K_\caN \prod_a J_{n_a}(u_a)
\label{eq:A-N}
\end{align}
\end{subequations}
are the wave number, initial phase and amplitude of the Rabi mode $\caN$
with 
\begin{align}
u_a &= \frac{\ctm\lambda_a}{k_a}.
\end{align}
Therefore, one expects that the flavor transformation of the neutrino
is enhanced when the Rabi resonance condition
\begin{align}
K_\caN = \wm
\label{eq:resonance-N}
\end{align}
is approximately met. 
We note that the resonance condition is
independent of the initial phases of the Rabi modes.

\section{Further discussion on Rabi resonances
\label{sec:discussion}}

\subsection{Amplitudes of the Rabi modes}
\label{sec:amp}
The physics prescription presented in Sec.~\ref{sec:multi-freq} seems
simple and appealing, but there remain a few questions that need to be
answered. First and foremost, there can exist many Fourier modes in a
realistic matter profile. If $\lambda_0(r)$ is a slowly varying
function of distance $r$ (as in most
realistic cases), at any given point one can almost always find some
or even many choices of $\caN$
with which the resonance condition in 
Eq.~\eqref{eq:resonance-N} is approximately satisfied. And yet Patton
et al found 
that only a few resonances were needed to account for the neutrino
flavor transformation through (at least some of)
the matter profiles in supernovae
\cite{Patton:2014lza}. They proposed that a parametric resonance 
is applicable only when the density scale height 
\begin{align}
h(r) = \lambda_0 \left|\frac{\rmd \lambda_0}{\rmd r}\right|^{-1}
\end{align}
is longer than the length scale of the Rabi transition, or
\begin{align}
\Omega h \gtrsim 1.
\end{align}
This criterion makes physical sense because we have assumed $\lambda_0$ to be
constant in Sec.~\ref{sec:multi-freq} which is approximately true on the
length scale of $h$.

Here we would like to point out that, even if many
harmonic parametric resonances may exist for a given oscillatory
matter profile,  only a finite number,
probably just a few, of them are relevant in a physical problem. The reason
is the following. The Rabi oscillation frequency is determined by the
Rabi mode $R$ that is (approximately) on resonance, i.e., $\Omega\approx A_R$.
Using Eqs.~\eqref{eq:J-asym} and \eqref{eq:A-N}  and identity
$J_{-n}(z)=(-1)^{n}J_{n}(z)$ we obtain
\begin{align}
A_R \sim \wm \tan(2\thetam) 
\sideset{}{'}\prod_a\left(\frac{\lambda_a}{k_a}\right)^{|n_a|}
\sideset{}{''}\prod_b   J_{n_b}(u_b)
\label{eq:A-approx}
\end{align}
where $\prod'$ includes all the ``regular'' Fourier modes with
$\lambda_a/k_a\ll 1$, and $\prod''$ includes the rest of the
Fourier modes. One expects that most of the Fourier modes are regular if the
fluctuation amplitude of the matter profile 
is small. 
We will call $|n_a|$ the ``order of contribution'' to the Rabi mode $R$ by the
$a$th Fourier mode. A Fourier mode is ``standby'' if $n_a=0$ and
``participating'' otherwise.%
\footnote{It was pointed out by Patton et al that
  a standby Fourier mode $b$ can kill the parametric resonance if $u_b$
  happens to be 
 a root of $J_0(z)$ \cite{Patton:2013dba}. This can happen only if the
 standby Fourier mode is not a regular mode.
\label{fn:lw}}
From Eq.~\eqref{eq:A-approx} one sees that, there can be
only a few participating, regular Fourier modes and the order of
contribution of each of these Fourier modes must be small. Otherwise,
the amplitude $A_R$ of
the Rabi mode will be too small to be relevant.
If, however, the amplitude of a Fourier mode $b$ is so large
or its wavelength is so long (but is still shorter than $h$ or
the physical size of the system) that
$\lambda_b/k_b\gtrsim 1$, then,
according to Eq.~\eqref{eq:J-asym}, it can contribute to the Rabi mode
up to the order of $|n_b|\lesssim (\lambda_b/k_b)^2$ or the amplitude of
the Rabi mode will be again too small to be relevant. 

The above constraints on the contribution orders of the Fourier modes put a
stringent limit on the number of the on-resonance Rabi modes that one needs to
consider in a real physical problem.

\subsection{Interference between Rabi modes}
\label{sec:interference}

The Rabi formula in Eq.~\eqref{eq:P} was derived
assuming that there exists only one Rabi mode. In
Refs.~\cite{Kneller:2012id,Patton:2013dba,Patton:2014lza} the
rotating wave approximation was employed which is equivalent to ignoring
all the Rabi modes that are off-resonance.
However, under certain conditions the
rotating wave approximation may fail, and
off-resonance Rabi modes can interfere with the on-resonance mode 
 as we will show below.%
\footnote{The interference between Rabi modes discussed here is
  different than the suppression of the parametric resonance by
  certain long-wavelength Fourier modes which was discussed in
  Ref.~\cite{Patton:2013dba} (see also footnote~\ref{fn:lw}) and the
  three-flavor effect discussed in Ref.~\cite{Yang:2015oya}.}

We first consider a Rabi system with an on-resonance mode $R$ 
and an off-resonance mode $O$. The Hamiltonian of the system is the
same as that in Eq.~\eqref{eq:H-rot} except with $\caN=R$ and $O$
only. We define a new basis  
\begin{align}
\begin{bmatrix}
\ket{\nu'_1} \\ \ket{\nu'_2}
\end{bmatrix}
=\begin{bmatrix}
\cos\Theta & -\sin\Theta \\
\sin\Theta & \cos\Theta
\end{bmatrix}
\begin{bmatrix}
e^{-\rmi(K_O r+\Phi_O)/2}\ket{\nutm_1} \\ 
e^{\rmi(K_O r+\Phi_O)/2}\ket{\nutm_2}
\end{bmatrix},
\label{eq:rot2}
\end{align}
where
\begin{align}
\Theta &= \frac{1}{2} \arctan\left(\frac{A_O}{\wm-K_O}\right).
\end{align}
The Hamiltonian in this new basis is
\begin{widetext}
\begin{align}
\sfH' = \frac{1}{2} \begin{bmatrix}
-\wm'-A_R\cos(\Upsilon(r))\sin(2\Theta)  &
A_R(e^{\rmi \Upsilon(r)}\cos^2\Theta  - e^{-\rmi \Upsilon(r)}\sin^2\Theta) \\
A_R(e^{-\rmi \Upsilon(r)}\cos^2\Theta  - e^{\rmi \Upsilon(r)}\sin^2\Theta) &
\wm' + A_R\cos(\Upsilon(r))\sin(2\Theta) 
\end{bmatrix}
\label{eq:H-rot2}
\end{align}
\end{widetext}
where
\begin{align}
\wm' &= \text{sgn}(\wm-K_O)\sqrt{(\wm-K_O)^2+A_O^2}
\end{align}
and
\begin{align}
\Upsilon(r) = (K_R + K_O) r + \Phi_R + \Phi_O.
\end{align}
Because $A_R$ is small, we will keep only the off-diagonal
oscillatory terms in 
$\sfH'$ that are approximately on resonance so that
\begin{align}
\sfH' \approx \frac{1}{2}\begin{bmatrix}
-\wm' & A_R e^{\rmi\Upsilon(r)} \\
A_R e^{-\rmi\Upsilon(r)} & \wm'
\end{bmatrix}.
\label{eq:H-p}
\end{align}
This is exactly the Hamiltonian for a single-mode Rabi system.
Therefore,  a resonance occurs when
\begin{align}
K_R = \wm' = \wm + \Delta \wm
\label{eq:resonance-int}
\end{align}
where
\begin{align}
\Delta\wm = \wm'-(\wm-K_O) \approx \frac{A_O^2/2}{\wm-K_O}.
\label{eq:dw}
\end{align}
Comparing Eqs.~\eqref{eq:resonance-N} and \eqref{eq:resonance-int} one
sees that the resonance frequency is shifted by
$\Delta\wm$ because of the off-resonance mode.
This shift of the resonance frequency due to the off-resonance
Rabi modes is known as the ac Stark effect (see, e.g.,
Ref.~\cite{Cohen-Tannoudji:1996}).%
\footnote{The resonance shift due to the $n=-1$ mode of the
  Rabi Hamiltonian in Eq.~\eqref{eq:H-rabi} is known as the Bloch-Siegert
  shift \cite{Bloch:1940,Shirley:1965}. The shifts due to the
  other Fourier/Rabi modes 
  can be considered as the generalized Bloch-Siegert
  shift \cite{Tuorila:2010}.}
The new relative
detuning of the Rabi system is
\begin{align}
D'_R = \left|\frac{K_R-(\wm+\Delta\wm)}{A_R}\right|.
\label{eq:Dp}
\end{align}
The off-resonance mode will have a significant impact on the resonance if
the change of the relative detuning
\begin{align}
\Delta D_R = |D'_R-D_R| = \left|\frac{\Delta\wm}{A_R}\right|
\label{eq:dD}
\end{align}
is of order 1 or larger, or, equivalently,
\begin{align}
  |A_O| \gtrsim \sqrt{2|A_R(\wm-K_O)|} \sim \sqrt{|A_R| \wm}.
  \label{eq:crit-Ao}
\end{align}
This explains why the off-resonance Rabi modes can be ignored in the
case with a single Fourier mode (see Fig.~\ref{fig:single-mode}). When
the $n=1$ mode is almost on resonance, the $n=-1$ mode does not satisfies
the criterion in Eq.~\eqref{eq:crit-Ao} because
$A_{-1}=A_1\ll\wm$. The Rabi modes with $|n|>1$ have even smaller
amplitudes than the $n=-1$ mode.

We note that the Rabi system with
two Rabi modes describes a magnetic dipole in the presence of three
magnetic fields: $\bfB_0$ 
in the $z$ direction which corresponds to the diagonal elements of the
Hamiltonian $\sfH$, and $\bfB_R$ and $\bfB_O$ which rotate 
in the $x$-$y$ plane with
different angular frequencies $K_R$ and $K_O$
and which correspond the two Rabi modes in the
off-diagonal element of $\sfH$. The essence of 
Eqs.~\eqref{eq:rot2} and \eqref{eq:H-rot2} is to transform the equation
of motion from the static frame to the reference frame
which co-rotates with  $\bfB_O$. In this rotating frame one
has only one rotating field 
$\bfB_R'$ and one static field $\bfB_0'+\bfB_O'$, where the primes
indicate the quantities in the rotating frame.
The static field $\bfB_0'+\bfB_O'$ is titled
away from the $z$ axis by an angle $2\Theta$.%
\footnote{The transformation in Eq.~\eqref{eq:rot2} also rotates the
  system so that the static field $\bfB_0'+\bfB_O'$ is in the $z$
  direction.}. 
Because we consider the scenarios
where all the rotating fields have amplitudes much smaller than $|\bfB_0|$,
$\Theta$ is small and can be ignored.
Therefore, the system in
the co-rotating frame corresponds to a Rabi system with only one Rabi
mode $\bfB_R'$ the properties of which are given by Eqs.~\eqref{eq:P},
\eqref{eq:D} and \eqref{eq:Omega}. The
interference effect due to the off-resonance Rabi mode $\bfB_O$ is
manifested in the change of the magnitude of the static field
$|\bfB_0|\rightarrow|\bfB_0'+\bfB_O'|$.

For a Rabi system with one on-resonance Rabi mode and two
off-resonance Rabi modes all of which have small amplitudes, one can
transform the equation of motion to the reference frame which
co-rotates with one of the off-resonance mode. In this reference frame
there are only two Rabi modes and the energy gap $\wm$ changes to
$\wm'$. One can then applies the results of the two-mode Rabi system
that we discussed above.
In general, for a Rabi system with $N$ small-amplitude Rabi modes, one can
always go to the reference frame that co-rotates with one of the
off-resonance Rabi mode. In this co-rotating frame the number of Rabi
modes is reduced by one, and one can apply the results of the Rabi
system with $N-1$ modes. Using the reduction procedure we find
that, for the scenario with one on-resonance Rabi mode and many
off-resonance modes,
Eq.~\eqref{eq:dw} is generalized to
\begin{align}
  \Delta\wm \approx \sum_O \frac{A_O^2/2}{\wm-K_O},
  \label{eq:dw-gen}
\end{align}
where the summation is carried over all the off-resonance Rabi modes. In
particular, if only a pair of off-resonance Rabi modes $O_\pm$ have large
enough amplitudes to affect the resonance, and if $A_{O_+}=A_{O_-}$ and
$K_{O_+}=-K_{O_-}$, we have
\begin{align}
  \Delta\wm \approx \frac{A_{O+}^2\wm}{\wm^2-K_{O_+}^2}.
\end{align}
The relative detuning of the multi-mode Rabi system  is still given by
Eq.~\eqref{eq:Dp}.

\begin{figure}[!htb]
  \includegraphics[width=\columnwidth]{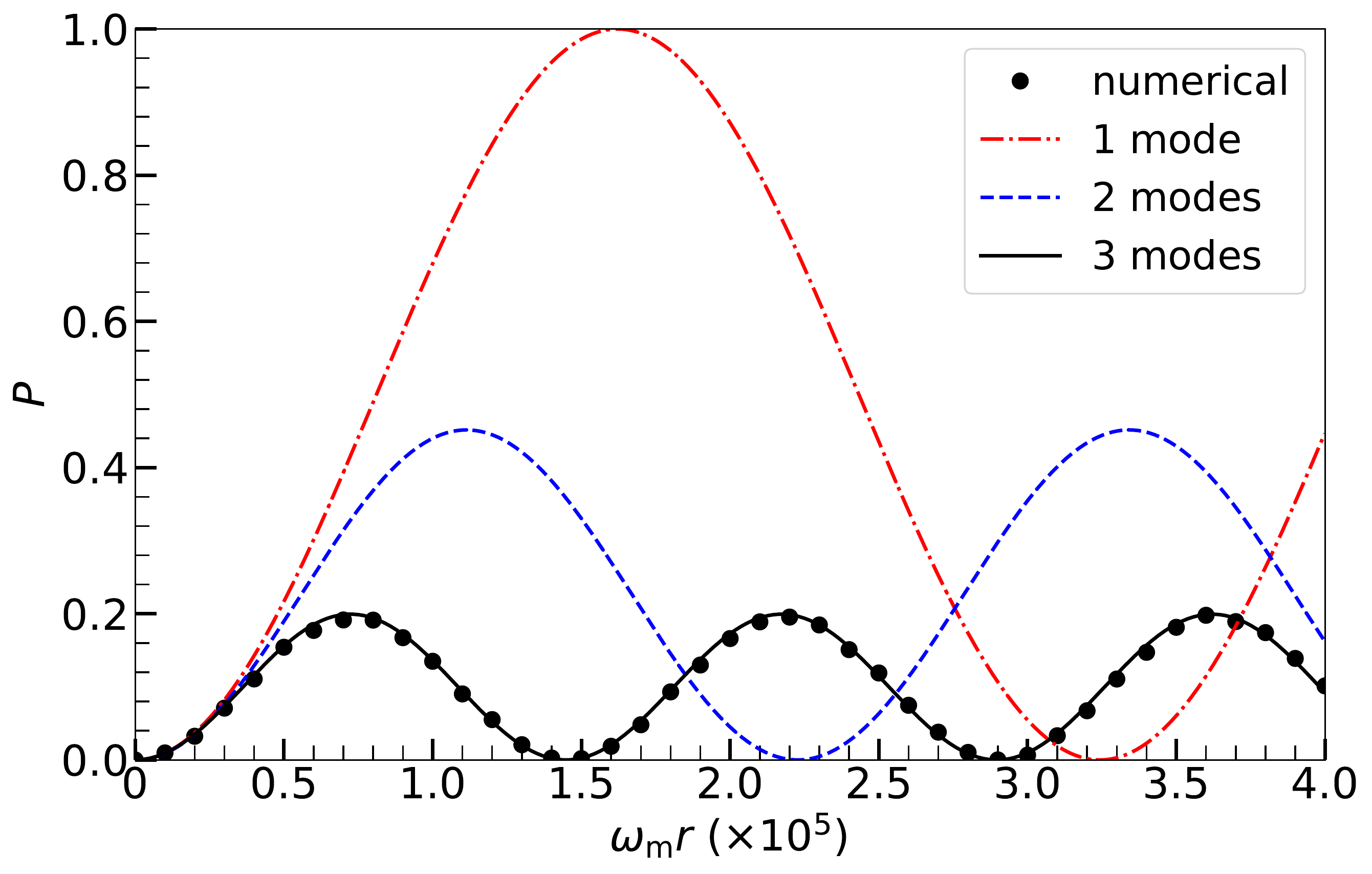}
  \caption{Similar to Fig.~\ref{fig:single-mode} but for a matter
    profile of the form $\lambda(r) = \lambda_0 +
    \lambda_1\cos(k_1 r) + \lambda_2\sin(k_2 r)$, where
    $\lambda_1/\wm=10^{-4}$, $k_1/\wm=1$,
    $\lambda_2/\wm=3.2\times10^{-2}$ and $k_2/\wm=0.1$.
    The filled circles represent the numerical solution to the Schr\"odinger
    equation, and the continuous curves represent the predictions by
    the Rabi formula when 1, 2 and 3 Rabi modes are included, respectively.} 
  \label{fig:int}
\end{figure}

As a concrete example 
we consider a matter profile of two Fourier modes:
\begin{align}
\lambda(r) = \lambda_0 + \lambda_1\cos(k_1 r) + \lambda_2\sin(k_2 r).
\end{align}
We choose $k_1=\wm$ so that the Rabi mode $R=\{1,0\}$ is exactly on
resonance. We choose the second Fourier mode to have a long wavelength
($k_2=0.1\wm$) and a relatively large amplitude
($\lambda_2=320\lambda_1=3.2\times10^{-2}\wm$). 
We compute the transition probability $P$ between $\ket{\nu_1^\mm}$
and $\ket{\nu_2^\mm}$ as a function of distance $r$ by solving the
Schr\"odinger equation numerically, and the result is shown 
in Fig.~\ref{fig:int}. As comparison 
we also show in the same figure the transition probabilities predicted
by the Rabi
formula when only the on-resonance Rabi mode $R=\{1,0\}$ is included,
both the $R$ mode and an off-resonance mode $O_+=\{0,1\}$ are
included, and the $R$
mode and two off-resonance modes $O_+$ and  $O_-=\{0,-1\}$ are
included, respectively. One can see that the numerical
solution agrees very well with the prediction based on the Rabi
formula when three Rabi modes $R$ and $O_\pm$ are included. One can
also see that the two 
long-wavelength, off-resonance Rabi modes $O_\pm$ combine to suppress
the Rabi transition.

\begin{figure}[!htb]
  \includegraphics[width=\columnwidth]{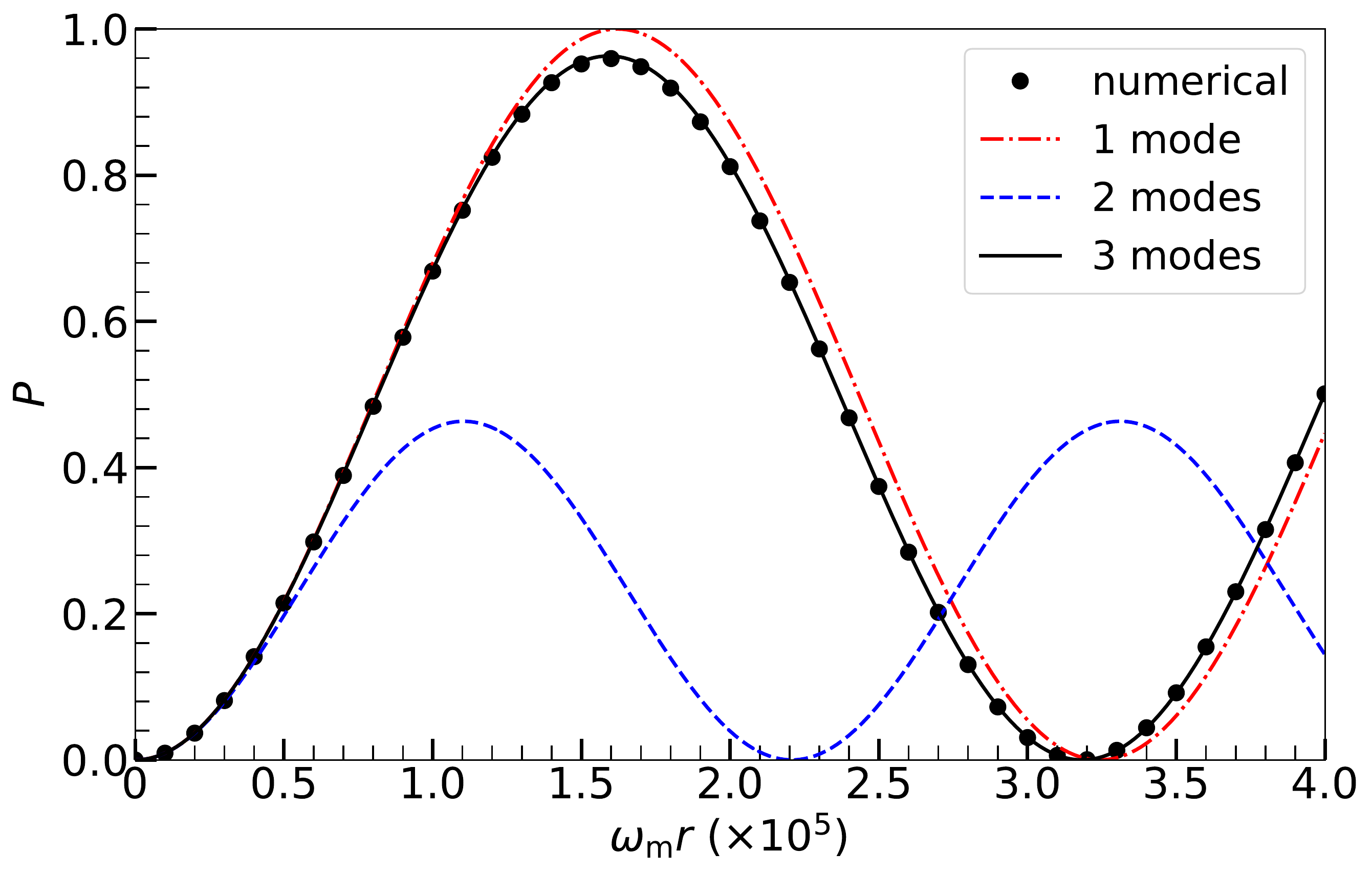}
  \caption{Same as Fig.~\ref{fig:int} but with $\lambda_2/\wm=0.1$ and
  $k_2/\wm=10$.} 
  \label{fig:int2}
\end{figure}

In Fig.~\ref{fig:int2} we demonstrate another case with the second
Fourier mode being a short-wavelength mode ($k_2=10\wm$ and
$\lambda_2=0.1\wm$). In this 
case, although each of the two off-resonance Rabi modes $O_\pm$ is
capable to suppress the Rabi transition by a large amount, the shifts
of the resonance frequency 
due to these two modes are in opposite directions [see Eq.~\eqref{eq:dw}]. As
a result, the suppression of the Rabi transition is not significant
in the actual system.

We note that, according to the discussion
in Sec.~\ref{sec:amp}, for a given on-resonance Rabi mode (which is
relevant to a physical system), Eq.~\eqref{eq:A-approx} implies that
only a finite 
number of off-resonance modes can satisfy the criterion in
Eq.~\eqref{eq:crit-Ao},
although an infinite number of Rabi modes exist in the system due to
the Jacobi-Anger expansion.
For small perturbations where the amplitudes of all Rabi modes are
much smaller than $\wm$, Eq.~\eqref{eq:crit-Ao}
demands that the amplitude of 
a single off-resonant Rabi mode must be significantly larger than that of the
on-resonance mode to affect the resonance. This can be true if the
on-resonant Rabi mode involves a few participating Fourier modes and,
therefore, has an amplitude much smaller than that of an off-resonant Rabi
mode with only one 
participating Fourier mode. Even for an on-resonant Rabi mode with
a single participating Fourier mode, the ac Stark shifts of many
off-resonance, long-wavelength Rabi modes can add up and
 change the resonant behavior according to
Eq.~\eqref{eq:dw-gen}.

\section{Conclusions
\label{sec:conclusions}}

We have shown
that the neutrino oscillation Hamiltonian with an oscillatory matter
profile can be treated as a Rabi system with an infinite number of
Rabi modes each with contributions from various Fourier modes of the
matter profile. Neutrino flavor 
conversion can be greatly enhanced if a Rabi mode is almost on
resonance. Although the existence of the harmonic parametric
resonances have already been shown in
Refs.~\cite{Kneller:2012id,Patton:2013dba}, our derivation adds more
intuitive understanding to this interesting phenomenon. 

We have shown that the number of the Fourier modes that participate
in a Rabi mode and their contribution orders cannot be too large or the
amplitude of the Rabi mode becomes too small to be relevant. As a
result, only a finite number of Rabi modes need to be considered for a
real physical problem.
We have also gone beyond the rotating wave approximation and studied the
interference between Rabi modes. This interference effect is different
than the suppression of the parametric resonance by 
certain long-wavelength Fourier modes discussed in
Ref.~\cite{Patton:2013dba}. It is also different than
the three-flavor effect discussed in Ref.~\cite{Yang:2015oya}.
We have shown that an
off-resonance Rabi mode can significantly change the
parametric resonance of neutrino flavor conversion if the amplitude of
the off-resonance 
mode is sufficiently large. 
We have derived an explicit criterion of
whether an off-resonance Rabi mode can affect the parametric
resonance. 
A Fourier mode in the matter
fluctuation always results in (an infinite number of) pairs of Rabi
modes. Each pair of these Rabi modes have the same amplitude but rotate
in the opposite directions. 
We found that the interference
effects due to a pair of such Rabi modes add up coherently if they have
long wavelengths, and they tend to
cancel each other if the wavelengths of the Rabi modes are short. As a
result, the Fourier modes with long wavelengths are much more likely to
affect the parametric resonance than the short-wavelength modes.

\acknowledgments{
We thank J.~Kneller, J.~Martin and K.~Patton for useful discussion.
This work was supported by the US DOE EPSCoR grant \#DE-SC0008142 and NP
grant \#DE-SC0017803 at UNM.}

\bibliographystyle{elsarticle-num}
\bibliography{paratran}

\begin{thebibliography}{10}
\expandafter\ifx\csname url\endcsname\relax
  \def\url#1{\texttt{#1}}\fi
\expandafter\ifx\csname urlprefix\endcsname\relax\def\urlprefix{URL }\fi
\expandafter\ifx\csname href\endcsname\relax
  \def\href#1#2{#2} \def\path#1{#1}\fi

\bibitem{Robertson:2012ib}
W.~C. Haxton, R.~G. Hamish~Robertson, A.~M. Serenelli, {Solar Neutrinos: Status
  and Prospects}, Ann. Rev. Astron. Astrophys. 51 (2013) 21--61.
\newblock \href {http://arxiv.org/abs/1208.5723} {\path{arXiv:1208.5723}},
  \href {http://dx.doi.org/10.1146/annurev-astro-081811-125539}
  {\path{doi:10.1146/annurev-astro-081811-125539}}.

\bibitem{Mirizzi:2015eza}
A.~Mirizzi, I.~Tamborra, H.-T. Janka, N.~Saviano, K.~Scholberg, R.~Bollig,
  L.~Hudepohl, S.~Chakraborty, {Supernova Neutrinos: Production, Oscillations
  and Detection}, Riv. Nuovo Cim. 39~(1-2) (2016) 1--112.
\newblock \href {http://arxiv.org/abs/1508.00785} {\path{arXiv:1508.00785}},
  \href {http://dx.doi.org/10.1393/ncr/i2016-10120-8}
  {\path{doi:10.1393/ncr/i2016-10120-8}}.

\bibitem{Wolfenstein:1977ue}
L.~Wolfenstein, Neutrino oscillations in matter, Phys. Rev. D17 (1978) 2369.

\bibitem{Mikheyev:1985aa}
S.~P. Mikheyev, A.~Y. Smirnov, Resonance ehancement of oscillations in matter
  and solar neutrino spectroscopy, Yad. Fiz. 42 (1985) 1441, [Sov. J. Nucl.
  Phys. 42, 913 (1985)].

\bibitem{Haxton:1990qb}
W.~C. Haxton, W.~M. Zhang, {Solar weak currents, neutrino oscillations and time
  variations}, Phys. Rev. D43 (1991) 2484--2494.
\newblock \href {http://dx.doi.org/10.1103/PhysRevD.43.2484}
  {\path{doi:10.1103/PhysRevD.43.2484}}.

\bibitem{Sawyer:1990tw}
R.~F. Sawyer, {Neutrino oscillations in inhomogeneous matter}, Phys. Rev. D42
  (1990) 3908--3917.
\newblock \href {http://dx.doi.org/10.1103/PhysRevD.42.3908}
  {\path{doi:10.1103/PhysRevD.42.3908}}.

\bibitem{Nunokawa:1996qu}
H.~Nunokawa, A.~Rossi, V.~B. Semikoz, J.~W.~F. Valle, {The Effect of random
  matter density perturbations on the MSW solution to the solar neutrino
  problem}, Nucl. Phys. B472 (1996) 495--517.
\newblock \href {http://arxiv.org/abs/hep-ph/9602307}
  {\path{arXiv:hep-ph/9602307}}, \href
  {http://dx.doi.org/10.1016/0550-3213(96)00236-2}
  {\path{doi:10.1016/0550-3213(96)00236-2}}.

\bibitem{Burgess:1996mz}
C.~P. Burgess, D.~Michaud, {Neutrino propagation in a fluctuating sun}, Annals
  Phys. 256 (1997) 1--38.
\newblock \href {http://arxiv.org/abs/hep-ph/9606295}
  {\path{arXiv:hep-ph/9606295}}, \href
  {http://dx.doi.org/10.1006/aphy.1996.5660}
  {\path{doi:10.1006/aphy.1996.5660}}.

\bibitem{Loreti:1995ae}
F.~N. Loreti, Y.~Z. Qian, G.~M. Fuller, A.~B. Balantekin, {Effects of random
  density fluctuations on matter enhanced neutrino flavor transitions in
  supernovae and implications for supernova dynamics and nucleosynthesis},
  Phys. Rev. D52 (1995) 6664--6670.
\newblock \href {http://arxiv.org/abs/astro-ph/9508106}
  {\path{arXiv:astro-ph/9508106}}, \href
  {http://dx.doi.org/10.1103/PhysRevD.52.6664}
  {\path{doi:10.1103/PhysRevD.52.6664}}.

\bibitem{Friedland:2006ta}
A.~Friedland, A.~Gruzinov, {Neutrino signatures of supernova turbulence}\href
  {http://arxiv.org/abs/astro-ph/0607244} {\path{arXiv:astro-ph/0607244}}.

\bibitem{Fogli:2006xy}
G.~L. Fogli, E.~Lisi, A.~Mirizzi, D.~Montanino, {Damping of supernova neutrino
  transitions in stochastic shock-wave density profiles}, JCAP 0606 (2006) 012.
\newblock \href {http://arxiv.org/abs/hep-ph/0603033}
  {\path{arXiv:hep-ph/0603033}}, \href
  {http://dx.doi.org/10.1088/1475-7516/2006/06/012}
  {\path{doi:10.1088/1475-7516/2006/06/012}}.

\bibitem{Kneller:2010sc}
J.~P. Kneller, C.~Volpe, {Turbulence effects on supernova neutrinos}, Phys.
  Rev. D82 (2010) 123004.
\newblock \href {http://arxiv.org/abs/1006.0913} {\path{arXiv:1006.0913}},
  \href {http://dx.doi.org/10.1103/PhysRevD.82.123004}
  {\path{doi:10.1103/PhysRevD.82.123004}}.

\bibitem{Choubey:2007ga}
S.~Choubey, N.~P. Harries, G.~G. Ross, {Turbulent supernova shock waves and the
  sterile neutrino signature in megaton water detectors}, Phys. Rev. D76 (2007)
  073013.
\newblock \href {http://arxiv.org/abs/hep-ph/0703092}
  {\path{arXiv:hep-ph/0703092}}, \href
  {http://dx.doi.org/10.1103/PhysRevD.76.073013}
  {\path{doi:10.1103/PhysRevD.76.073013}}.

\bibitem{Akhmedov:1999ty}
E.~K. Akhmedov, {Parametric resonance in neutrino oscillations in matter},
  Pramana 54 (2000) 47--63.
\newblock \href {http://arxiv.org/abs/hep-ph/9907435}
  {\path{arXiv:hep-ph/9907435}}, \href
  {http://dx.doi.org/10.1007/s12043-000-0006-4}
  {\path{doi:10.1007/s12043-000-0006-4}}.

\bibitem{Krastev:1989ix}
P.~I. Krastev, A.~{\relax Yu}. Smirnov, {Parametric Effects in Neutrino
  Oscillations}, Phys. Lett. B226 (1989) 341--346.
\newblock \href {http://dx.doi.org/10.1016/0370-2693(89)91206-9}
  {\path{doi:10.1016/0370-2693(89)91206-9}}.

\bibitem{Kneller:2012id}
J.~P. Kneller, G.~C. McLaughlin, K.~M. Patton, {Stimulated Neutrino
  Transformation in Supernovae}, J. Phys. G40 (2013) 055002.
\newblock \href {http://arxiv.org/abs/1202.0776} {\path{arXiv:1202.0776}},
  \href {http://dx.doi.org/10.1088/0954-3899/40/5/055002}
  {\path{doi:10.1088/0954-3899/40/5/055002}}.

\bibitem{Patton:2013dba}
K.~M. Patton, J.~P. Kneller, G.~C. McLaughlin, {Stimulated Neutrino
  Transformation Through Turbulence}, Phys. Rev. D89~(7) (2014) 073022.
\newblock \href {http://arxiv.org/abs/1310.5643} {\path{arXiv:1310.5643}},
  \href {http://dx.doi.org/10.1103/PhysRevD.89.073022}
  {\path{doi:10.1103/PhysRevD.89.073022}}.

\bibitem{Patton:2014lza}
K.~M. Patton, J.~P. Kneller, G.~C. McLaughlin, {Stimulated neutrino
  transformation through turbulence on a changing density profile and
  application to supernovae}, Phys. Rev. D91~(2) (2015) 025001.
\newblock \href {http://arxiv.org/abs/1407.7835} {\path{arXiv:1407.7835}},
  \href {http://dx.doi.org/10.1103/PhysRevD.91.025001}
  {\path{doi:10.1103/PhysRevD.91.025001}}.

\bibitem{Yang:2015oya}
Y.~Yang, J.~P. Kneller, {Neutrino Flavour Evolution Through Fluctuating
  Matter}, J. Phys. G45~(4) (2018) 045201.
\newblock \href {http://arxiv.org/abs/1510.01998} {\path{arXiv:1510.01998}},
  \href {http://dx.doi.org/10.1088/1361-6471/aab0c4}
  {\path{doi:10.1088/1361-6471/aab0c4}}.

\bibitem{Duan:2010bg}
H.~Duan, G.~M. Fuller, Y.-Z. Qian, {Collective Neutrino Oscillations}, Ann.
  Rev. Nucl. Part. Sci. 60 (2010) 569.
\newblock \href {http://arxiv.org/abs/1001.2799} {\path{arXiv:1001.2799}}.

\bibitem{Chakraborty:2016yeg}
S.~Chakraborty, R.~Hansen, I.~Izaguirre, G.~Raffelt, {Collective neutrino
  flavor conversion: Recent developments}, Nucl. Phys. B908 (2016) 366--381.
\newblock \href {http://arxiv.org/abs/1602.02766} {\path{arXiv:1602.02766}},
  \href {http://dx.doi.org/10.1016/j.nuclphysb.2016.02.012}
  {\path{doi:10.1016/j.nuclphysb.2016.02.012}}.

\bibitem{sakurai2011modern}
J.~Sakurai, J.~Napolitano, Modern Quantum Mechanics, 2nd Edition,
  Addison-Wesley, 2011.

\bibitem{Cohen-Tannoudji:1996}
C.~N. Cohen-Tannoudji, The autler-townes effect revisited, in: R.~Y. Chiao
  (Ed.), Amazing Light, Springer, New York, 1996, Ch.~11, pp. 109--123.

\bibitem{Bloch:1940}
F.~Bloch, A.~Siegert,
  \href{https://link.aps.org/doi/10.1103/PhysRev.57.522}{Magnetic resonance for
  nonrotating fields}, Phys. Rev. 57 (1940) 522--527.
\newblock \href {http://dx.doi.org/10.1103/PhysRev.57.522}
  {\path{doi:10.1103/PhysRev.57.522}}.
\newline\urlprefix\url{https://link.aps.org/doi/10.1103/PhysRev.57.522}

\bibitem{Shirley:1965}
J.~H. Shirley,
  \href{https://link.aps.org/doi/10.1103/PhysRev.138.B979}{{Solution of the
  Schrödinger equation with a Hamiltonian periodic in time}}, Phys. Rev. 138
  (1965) B979--B987.
\newblock \href {http://dx.doi.org/10.1103/PhysRev.138.B979}
  {\path{doi:10.1103/PhysRev.138.B979}}.
\newline\urlprefix\url{https://link.aps.org/doi/10.1103/PhysRev.138.B979}

\bibitem{Tuorila:2010}
J.~Tuorila, M.~Silveri, M.~Sillanp\"a\"a, E.~Thuneberg, Y.~Makhlin, P.~Hakonen,
  \href{http://link.aps.org/supplemental/10.1103/PhysRevLett.105.257003}{Stark
  effect and generalized bloch-siegert shift in a strongly driven two-level
  systemi: Supplementary information}, Phys. Rev. Lett. 105 (2010) 257003.
\newblock \href {http://dx.doi.org/10.1103/PhysRevLett.105.257003}
  {\path{doi:10.1103/PhysRevLett.105.257003}}.
\newline\urlprefix\url{http://link.aps.org/supplemental/10.1103/PhysRevLett.105.257003}

\end{thebibliography}

\end{document}